\documentclass{article}

\usepackage{authblk}
\usepackage{cancel}
\usepackage{array}
\usepackage{amsmath}
\usepackage{amssymb}
\usepackage{graphicx}
\usepackage{algorithm}
\usepackage{algpseudocode}
\usepackage[font=small]{caption}

\let\citeleft=(
\let\citeright=)

\newcommand{\argmin}[1]{\operatorname*{argmin}_{#1}}
\def\*#1{\mathbf{#1}}

\begin{document}

\pdfinfo{
   /Author (AUTHORS)
   /Title (TITLE)
}

\title{\vspace{-2cm}Implicit Representation of GRAPPA Kernels for Fast MRI Reconstruction}

\author[1]{Daniel Abraham}
\author[1]{Mark Nishimura}
\author[2]{Xiaozhi Cao}
\author[2]{Congyu Liao}
\author[1,2]{Kawin Setsompop}

\affil[1]{\small Stanford Electrical Engineering}
\affil[2]{\small Stanford Radiology}
\maketitle

\vfill

\clearpage

\section*{Abstract}
MRI data is acquired in Fourier space/k-space. Data acquisition is typically performed on a Cartesian grid in this space to enable the use of a fast Fourier transform algorithm to achieve fast and efficient reconstruction. However, it has been shown that for multiple applications, non-Cartesian data acquisition can improve the performance of MR imaging by providing fast and more efficient data acquisition, and improving motion robustness. Nonetheless, the image reconstruction process of non-Cartesian data is more involved and can be time-consuming, even through the use of efficient algorithms such as non-uniform FFT (NUFFT). Reconstruction complexity is further exacerbated when imaging in the presence of field imperfections. This work (implicit GROG) provides an efficient approach to transform the field corrupted non-Cartesian data into clean Cartesian data, to achieve simpler and faster reconstruction which should help enable non-Cartesian data sampling to be performed more widely in MRI. 

\noindent
\textbf{Purpose}: 
Using non-Cartesian acquisition schemes allows for faster scanning, motion robustness, and more efficient use of the gradient coils. However, these non-Cartesian acquisitions are sensitive to field imperfections and hence will often result in a complex and costly reconstruction scheme. Implicit GROG is proposed to assist in reducing the reconstruction complexity.

\noindent
\textbf{Theory and Methods}:
The two main causes of costly reconstructions in MRI are off-grid (non-Cartesian) sampling and field correction techniques such as time segmentation. Inspired by GROG, implicit GROG tackles the off-grid sampling issue by estimating gridded k-space samples from non-Cartesian samples, enabling the use of the FFT for reconstruction. Unlike GROG, this process is generalized for multiple input non-Cartesian samples to allow for a more accurate estimation of gridded samples. The issue of field correction is handheld by modeling phase accrued by field imperfections as linear combination of MRI Fourier-weighted coil sensitivity maps. This allows for a partial field correction to be performed on the raw data using the implicit GROG framework, resulting in a dramatic reduction in time segments in reconstruction.  

\noindent
\textbf{Results}: 
In simulation we show that implicit GROG results in higher quality reconstructions in comparison to GROG with a lower noise amplification. Simulation results also show a 3-4X reduction in time segments needed to correct smooth field imperfections. Both claims are validated in-vivo.

\noindent
\textbf{Conclusion}: 
Implicit GROG is an effective framework for estimating clean Cartesian data from field imperfection corrupted non-Cartesian data. This enables clinically relevant reconstruction times for non-Cartesian sequences, while significantly reducing the memory and processing requirements.
  
\noindent
\textbf{Keywords}: 
Machine Learning, GRAPPA, SENSE

\clearpage

\section*{Introduction}
\label{sec:introduction}
At the time of this paper, a vast majority of clinical MR scans acquire data on a Cartesian grid. Cartesian acquisitions are advantageous due to their acquisition and reconstruction simplicity. Since Cartesian acquisitions acquire samples on a Cartesian grid, image reconstruction can be performed in an efficient way using Fast Fourier Transforms (FFTs). However, Cartesian acquisitions can often lead to longer scan times, due to the low encoding efficiency introduced by line-by-line scanning. Echo Planar Imaging (EPI) and multi-shot EPI techniques are solutions which improve the duty cycle of Cartesian-like trajectories, while retaining the efficient reconstruction benefits. Nonetheless, EPI trajectories still only utilize a single effective gradient coil for image encoding, limiting the overall encoding efficiency.

Non-Cartesian trajectories are able to remedy many of the short comings of Cartesian trajectories. Spiral trajectories enable 2D and 3D rapid imaging applications, allow for much shorter TEs in diffusion imaging, and are the standard choice for MRF acquisitions. Radial trajectories exhibit desirable undersampling artifacts and are extremely robust to motion. Both of these trajectories have been used for body imaging \cite{feng2016xd}, functional MRI \cite{Wiesinger2018}, and quantitative imaging. However, these trajectories have not become a standard choice in clinical settings, due to a multitude of reasons. All non-Cartesian acquisitions will require a non-Cartesian image reconstruction, which usually results in much longer reconstruction times. Additionally, some non-Cartesian acquisitions, such as spirals, can have a higher sensitivity to field imperfections, requiring off-resonance correction techniques (such as time segmentation) which further burden reconstruction times. Long reconstruction times can be an issue in clinical settings, since clinical scans often rely on being able to observe the reconstructed volumes shortly after data acquisition is complete to determine the need for a re-scan. Thus, it is important to have both the improved acquisition efficiency that Non-Cartesian trajectories offer, as well as the rapid reconstructions that Cartesian trajectories offer. 

Non-Uniform Fast Fourier Transforms (NUFFTs) \cite{fessler2003nonuniform} provide a significant computational improvement in comparison to the DFT approach for computing Non-Cartesian k-space samples from an image. The NUFFT algorithm first applies an over-sampled spatial FFT to the image to achieve the Cartesian k-space samples, and then uses a Kaiser Bessel kernel to interpolate the non-Cartesian samples from the gridded Cartesian ones. While NUFFTs have significantly improved MRI reconstruction, they are still considered a substantial bottleneck in modern Non-Cartesian reconstruction pipelines. In most model based Non-Cartesian reconstruction frameworks, the NUFFT and it's adjoint operator must be applied iteratively. To remedy the costly iterative NUFFT applications, many have used a Toeplitz \cite{baron2018rapid} formulation to speed up the successive applications of the NUFFT and it's adjoint operation. The Toeplitz formulation removes the need for a Kaiser Bessel interpolation, and instead requires the use of a $2\times$ oversampled FFT and storing the large Toeplitz Point Spread Function (PSF). While the Toeplitz formulation provides a significant improvement, the computation time of the $2\times$ oversampled FFTs and Toeplitz PSF memory requirements can become problematic for high dimensional and high resolution MR volumes. 

In addition to the time penalty incurred by reconstructing non-Cartesian data, correcting for field imperfections in reconstruction has an independent multiplicative penalty on reconstruction time. Almost all field correction techniques require splitting the accrued phase into a sum of spatial basis functions multiplied by temporal basis functions \cite{fessler2005toeplitz}\cite{man1997multifrequency}\cite{sutton2003fast}. As shown in \cite{fessler2005toeplitz}, time segmentation, a popular choice for the spatial and temporal basis functions, results in a nearly optimal field correction, and thus we will focus on time segmentation for the remainder of this work. If there are $L$ such time segments, then the reconstruction penalty is approximately $L$ fold. Furthermore, using the Toeplitz formulation with these time segmented reconstructions requires storing $L$ Toeplitz PSFs, which remains a significant bottleneck due to modern GPU memory limitations. 

An alternative Non-Cartesian reconstruction approach is to instead estimate Cartesian samples from Non-Cartesian samples using GRAPPA kernels, which is called GRAPPA Operator Gridding (GROG) \cite{seiberlich2007non}. Importantly, GROG estimates the Cartesian samples once before reconstruction, and the remaining reconstruction can be deployed on a Cartesian grid, which will be significantly faster than NUFFT based reconstructions since Kaiser Bessel interpolations are no longer needed. Furthermore reconstructing Cartesian data does not require the large Toeplitz PSFs, which helps address GPU memory limitations. Applying GRAPPA to non-Cartesian trajectories requires the training, storage, and application of a vast amount GRAPPA kernels, which can become a computational burden. To remedy this problem, GROG trains only a few GRAPPA kernels, one for each coordinate axis, and the applies them sequentially on a Non-Cartesian sample to estimate a Cartesian sample. Following training, gridding is performed on novel gridding geometries by using fractional powers of the trained coordinate GRAPPA kernels. GROG is fast and in certain cases can provide reconstruction results that are similar to the gold standard iterative NUFFT approaches. Unfortunately, GROG suffers from non-negligible errors in the estimated cartesian data points. These errors arise from the restricted use of single input (source) single output (target) GRAPPA kernels, which in some cases are not expressive enough to capture the full redundancy along the coil dimension. 

Implicit GROG is proposed to overcome the shortcomings of GROG, and provide a new way of partially correcting for field imperfections. In order to improve GROG's shortcomings, we propose to instead estimate a unique GRAPPA kernel for each gridding geometry, and input multiple Non-Cartesian source samples to estimate the single target Cartesian sample. Using multiple source points helps reduce the estimation errors, and makes better of use of the coil redundancy. In addition to this, implicit GROG estimates target Cartesian samples with reduced phase due to field imperfections. This is possible for field imperfections of low spatial order, as they can be modeled by Fourier weighted MRI coil sensitivity maps.

\section*{Theory}
\label{sec:theory}
Throughout the theory section, we will use the following notation 
\begin{center}
\begin{tabular}{ |c|c| } 
 \hline
 $d$ & number of spatial dimensions, typically 2 or 3  \\  
 \hline
 $C$ & number of MRI coil sensitivity maps \\  
 \hline
 $N^d$ & number of voxels (assuming $N$ points along each spatial dimension)\\  
 \hline
 $M$ & number of k-space samples \\  
 \hline
 $\mathbf{r}$ & spatial signal domain coordinate vector in $\mathbb{R}^d$ \\  
 \hline
 $\mathbf{k}$ & spatial Fourier domain coordinate vector in $\mathbb{R}^d$ \\  
 \hline
 $m(\mathbf{r})$ & magnetization volume $m :\mathbb{R}^d \rightarrow \mathbb{C}$ \\  \hline
 $\mathbf{s}(\mathbf{r})$ & MRI coil sensitivity maps $\mathbf{s} :\mathbb{R}^d \rightarrow \mathbb{C}^{C}$ \\  
 \hline
 $\phi(\mathbf{r}, t)$ & phase accrual due to field imperfections $\phi : \mathbb{R}^{d + 1} \rightarrow \mathbb{R}$ \\  
 \hline
 $\mathbf{M}(\mathbf{k}, t)$ & Multi-channel Fourier data with phase accrual $\mathbf{M} : \mathbb{R}^{d + 1} \rightarrow \mathbb{C}^{C}$ \\  
 \hline
 $\mathbf{b}(t)$ & raw multi-channel MRI data $\mathbf{b} :\mathbb{R} \rightarrow \mathbb{C}^{C}$ \\ 
 \hline
\end{tabular}
\end{center}

\subsection*{Non-Cartesian Reconstruction in the Presence of Field Imperfections}
We first relate the multi-channel Fourier transform of $m(\mathbf{r})$ at some Fourier coordinate $\mathbf{k}$ after $t$ seconds of phase accrual as 
\begin{equation}
\label{signal}
    \mathbf{M}(\mathbf{k}, t) = \int_\mathbf{r} m(\mathbf{r}) \mathbf{s}(\mathbf{r}) e^{-j2\pi \mathbf{k} \cdot \mathbf{r}} e^{-j2\pi \phi(\mathbf{r}, t)} d\mathbf{r}.
\end{equation}
With the formula above in mind, we can define the MRI forward model over some k-space trajectory $\mathbf{k}(t)$ as 
\begin{equation}
\label{forward}
    \mathbf{b}(t) = \mathbf{M}(\mathbf{k}(t), t)
\end{equation}
and the adjoint operator as 
\begin{equation}
\label{adjoint}
    \hat{m}(\mathbf{r}) = \bar{\mathbf{s}}(\mathbf{r}) \cdot \int_t \mathbf{b}(t) e^{j 2\pi \mathbf{k}(t) \cdot \mathbf{r}} e^{j 2\pi \phi(\mathbf{r}, t)} dt
\end{equation}

Most MRI reconstruction procedures will apply the normal operator in an iterative manner, which is simply the forward operation (\ref{forward}) followed by the adjoint operation (\ref{adjoint}). Thus, speeding up the evaluation of the forward model directly improves reconstruction times. 

In order for the forward model (\ref{forward}) to be more computationally efficient, the NUFFT algorithm is used for handling non-Cartesian trajectories $\mathbf{k}(t) \notin \mathbb{Z}^d$. This process involves an $\alpha$ fold oversampled FFT operation, followed by an interpolation over the entire $M$ point trajectory with a Kaiser Bessel kernel of width $W$. This however cannot directly be used when there exists phase accrual that contains at least second order spatial terms. Previous works \cite{fessler2005toeplitz}\cite{sutton2003fast}\cite{man1997multifrequency} suggest a low-rank splitting of the problematic spatio-temporal phase accrual term $e^{-j2\pi \phi(\mathbf{r}, t)}$ into $L$ spatial and temporal basis functions. We will use time-segmentation as the choice for the spatio-temporal splitting, which is shown to perform similarly to the optimal SVD decomposition \cite{fessler2005toeplitz} while being much more computationally efficient. The time segmentation decomposition is
\begin{equation}
    \label{field}
    e^{-j2\pi \phi(\mathbf{r}, t)} \approx \sum_{l=1}^L e^{-j 2\pi \phi(\mathbf{r}, t_l)} h_l(t),
\end{equation}
where $t_l$ are the time segment centers. Using this decomposition, there will be approximately $L$ multi-channel NUFFT evaluations for one full forward model evaluation. Combining the effects of both NUFFTs and modeling of field imperfections results in the following runtime for a forward (and adjoint) model evaluation:
\begin{equation}
    \label{runtime}
    O\bigg(L\cdot(\alpha^d N^d \log(\alpha N) + W^d M)\bigg) 
\end{equation}
where $1 < \alpha < 2$ is the NUFFT oversampling factor and $W$ is the kaiser bessel kernel width. As shown by \cite{fessler2005toeplitz}\cite{baron2018rapid}, the Toeplitz formulation can be used to speed up evaluations of the normal operator, which removes the dependence on the trajectory. This new formulation will have a computational complexity of  
\begin{equation}
    \label{normal}
    O\bigg(L \cdot 2^d N^d \log(2 N)\bigg) 
\end{equation}

However, using the Toeplitz formulation has a massive memory footprint due to the need to compute and store $L$ Toeplitz PSFs of size $(2N)^d$. This problem is further exacerbated in subspace reconstructions, which will require $L \cdot K^2$ Toeplitz PSFs, where $K$ is the number of subspace basis coefficients. For these reasons, and memory limitations on modern GPUs, we turn to methods that speed up the forward model evaluation without relying on the Toeplitz structure.

\subsection*{GROG}
Much of the prior parallel imaging works, such as GRAPPA \cite{griswold2002generalized}, have justified the use of linear functions, referred to as GRAPPA kernels, to estimate novel k-space samples from acquired ones. GRAPPA kernels estimate a single multi-channel k-space sample, typically referred to as the target point, from other neighboring multi-channel k-space samples, which are referred to as source points. A GRAPPA kernel is a function of the coil sensitivity maps $\mathbf{s}(\mathbf{r})$, and of the source point coordinates relative to the target point coordinate (which we call a gridding orientation).

GROG proposes to use a unique GRAPPA kernel for each coordinate axis, and then applies these kernels sequentially to estimate Cartesian samples from Non-Cartesian samples. In the GROG framework, GRAPPA kernels are trained to estimate target points which are $\Delta k = 1$ distance away from the source point along a single coordinate axis. All other distances can be modeled as fractional powers of the coordinate GRAPPA kernel. Let us define some multi-channel target sample $\mathbf{t}$ and a muilti-channel source sample $\mathbf{s}$ located at some position $(\delta_x, \delta_y, \delta_z)$ relative to $\mathbf{t}$. GROG estimates $\mathbf{t}$ from $\mathbf{s}$ using the following formulation
\begin{equation}
\label{grog}
    \mathbf{G}^{\delta_x} \mathbf{G}^{\delta_y} \mathbf{G}^{\delta_z} \mathbf{s} = \mathbf{t}
\end{equation}
where $\mathbf{G}_x, \mathbf{G}_y, \mathbf{G}_z$ are the trained GRAPPA kernels that estimate a point $\Delta k = 1$ away on each coordinate axis. This is illustrated in Figure \ref{fig:explain}a.

After applying GROG to the raw non-Cartesian MRI data $\mathbf{b}(t)$, a new set of Cartesian samples $\mathbf{b}_\text{grid}(t)$ and Cartesian k-space trajectory $\mathbf{k}_\text{grid}(t) \in \mathbb{Z}^d$ is formed. Since the new data is on a Cartesian grid, the Kaiser Bessel interpolation step is no longer necessary, leading to the new improved forward model runtime of 
\begin{equation}
    \label{grog_runtime}
    O\bigg( N^d \log(N)\bigg)
\end{equation}
Note that there is technically still a minor dependence on $M$, since after applying an FFT one must select the relevant grid samples. We choose to ignore this dependence as it negligible compared to the larger FFT operation.

While GROG provides a substantial improvement in reconstruction times, it has two main drawbacks. First, using just a single source point limits the GRAPPA kernels ability to estimate the target point, leading to artifacts and noise amplification. Furthermore, the GROG framework does not have functionality for correcting non-Fourier phase due to field imperfections. With these drawbacks in mind, we present a more general and flexible formulation of GROG.

\subsection*{Implicit GROG: Basic non-Cartesian}
\begin{figure}
    \centering
    \includegraphics[width=1\textwidth]{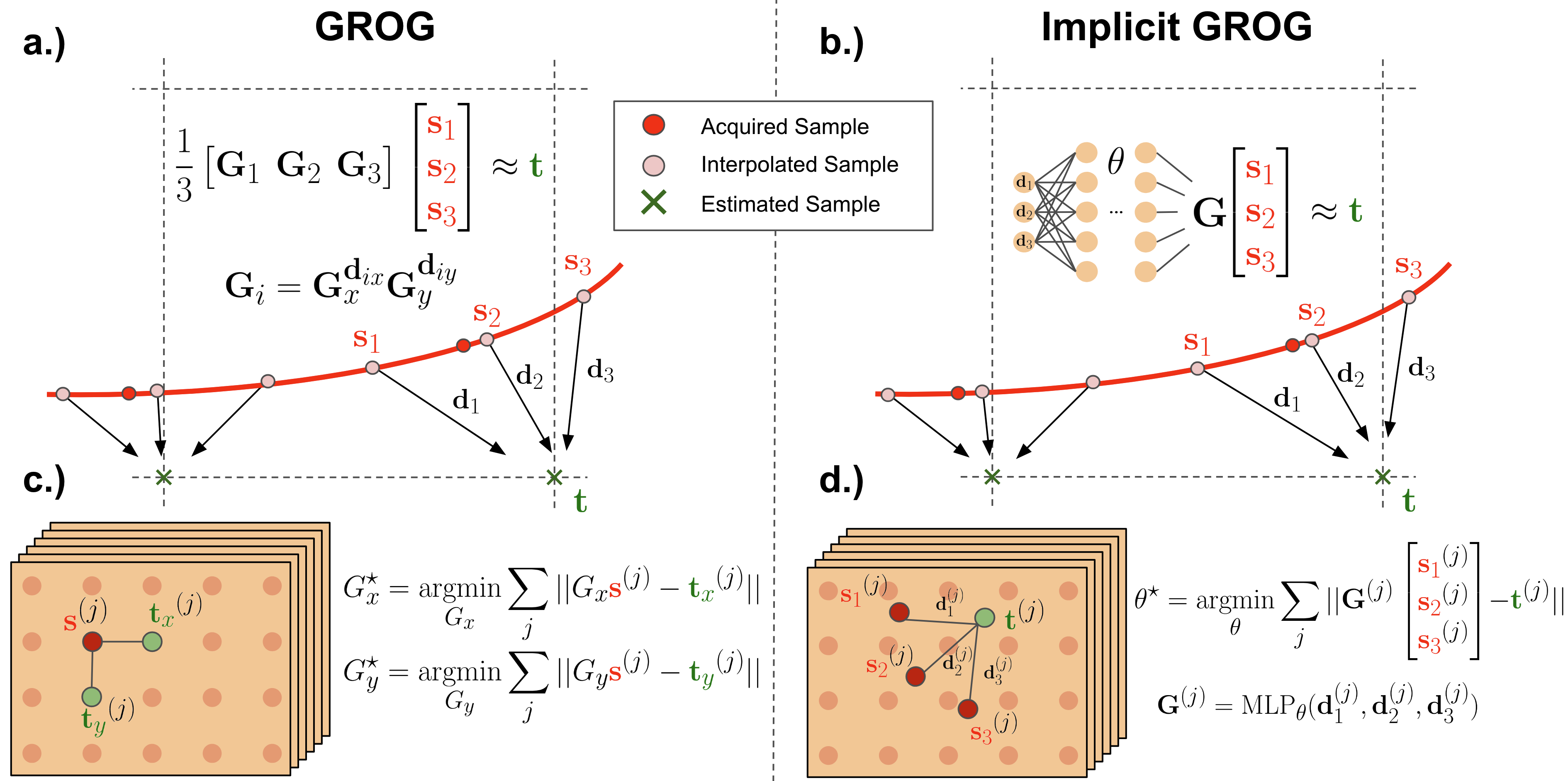}
    \caption{a.) The original GROG technique uses a single Non-Cartesian source point to estimate the target Cartesian sample. b.) implicit GROG instead uses multiple Non-Cartesian source points to estimate the target Cartesian sample c.) GROG trains coordinate GRAPPA kernels over the calibration signal, and sequentially apply fractional powers of these kernels for novel gridding orientations. d.) implicit GROG trains an implicit neural representation over the calibration signal and queries the model for novel gridding orientations}
    \label{fig:explain}
\end{figure}

Our method, implicit GROG, differs from GROG by instead using many non-Cartesian source points $(\mathbf{s}_1 \cdots \mathbf{s}_{D})$ to estimate the single target (Cartesian) point $\mathbf{t}$. The relationship between the source and target points is now 
\begin{equation}
\label{egrog}
    \mathbf{G} \begin{bmatrix} \mathbf{s}_1 & \cdots & \mathbf{s}_{D} \end{bmatrix}^\top = \mathbf{t}
\end{equation}
where $\mathbf{G}$ is the GRAPPA kernel which estimates a single target point from $D$ source points. Note that the source points are chosen to be along the readout direction, as illustrated in Figure \ref{fig:explain}b. This is done for simplicity and generality, although it is not strictly necessary.

The main challenge to this approach is that a unique GRAPPA kernel $\mathbf{G}$ is needed for each sample, since (in the worst case) each sample can posses a unique gridding orientation and image phase due to field imperfections. Naively estimating all of these kernels, even when using the fast GRAPPA estimation technique \cite{luo2019grappa}, is extremely computationally demanding. Training coordinate kernels as in GROG will not work here, since multiple source points are being used. To remedy this problem, the inherent correlation between unique GRAPPA kernels is exploited to train a lower dimensional representation of the space of all GRAPPA kernels needed for gridding. Inspired by recent success in computer vision tasks \cite{mildenhall2021nerf}, a Multi-Layer Perceptron (MLP) is used to \textit{implicitly} represent this low dimensional space of GRAPPA kernels. These MLPs offer a memory efficient and fast method of querying the required GRAPPA kernels, and have significantly lower training times in comparison to training the collection of GRAPPA kernels independently. 

 Let us define an orientation by the positions $\mathbf{d}_i$ of the source points $\mathbf{s}_i$ relative to the target point $\mathbf{t}$'s position. The implicit MLP formulation can be written as:
\begin{equation}
\label{mlp}
    \mathbf{G} = f_\theta(\mathbf{d}_1, \cdots, \mathbf{d}_{D})
\end{equation}
where $f_\theta$ is an MLP with parameters $\theta$. This MLP is a compressed representation of the collection of GRAPPA kernels needed to perform gridding. Training this compressed representation allows for a significant reduction in training time and memory in comparison to naively training each unique GRAPPA kernel. We formulate the training procedure as a stochastic minimization problem:
\begin{align}
\label{train}
    \theta^\star = \argmin{\theta} &\sum_j \mathcal{L}(\mathbf{G}^{(j)} \begin{bmatrix}
        \mathbf{s}_1^{(j)} & \cdots & \mathbf{s}_{D}^{(j)}
    \end{bmatrix}^\top,  \mathbf{t}^{(j)}) + \lambda ||\mathbf{G}^{(j)}||_F^2 \\
    \mathbf{G}^{(j)} &= f_\theta(\mathbf{d}_1^{(j)}, \cdots, \mathbf{d}_{D}^{(j)}) \nonumber
\end{align}
$\mathcal{L}$ is the loss function and $\lambda$ is a tikhonov regularization hyper-paremeter controlling the tradeoff between noise amplification and bias in the trained GRAPPA kernels. The MLP $f_\theta$ is trained on the calibration data, and is then applied on the non-Cartesian data to estimate the closest Cartesian data points. Since the calibration data points are usually acquired before each scan, the training of $f_\theta$ can be seen as a completely self-supervised process.

\subsection*{Implicit GROG Field Modeling}
In addition to correcting for non-Cartesian sampling, the implicit GROG framework can be modified to correct for field imperfections. In order to do this, let us first rewrite the arbitrary phase accrual as 
\begin{equation}
    \label{phase_decomp}
    \phi(\mathbf{r}, t) = \sum_{p=1}^P \phi_p(\mathbf{r}) \alpha_p(t) = \boldsymbol{\phi}(\mathbf{r}) \cdot \boldsymbol{\alpha}(t)
\end{equation}
where $\phi_p(\mathbf{r})$ are the spatial phase basis functions and $\alpha_p(t)$ are the temporal phase coefficients, and $\boldsymbol{\phi}(\mathbf{r})$, $\boldsymbol{\alpha}(t) \in \mathbb{R}^P$ are the vector representations. 
This decomposition is very useful, since it captures all types of field imperfections that occur in MRI in a compact way that will be useful for the MLP model. In the case of a main $B_0$ field inhomogeneity we have $P=1$, $\phi_1(\mathbf{r}) = B_0(\mathbf{r})$, and $\alpha_1(t) = t$. For trajectory dependent fields such as eddy currents, gradient non-linearities, and concomitant fields, the spatial phase basis functions can be spherical harmonics. 

In order to determine the time segmentation spatial basis functions, one can perform a clustering procedure directly on the discretized temporal coefficients $\begin{bmatrix} \boldsymbol{\alpha}(t_i) \end{bmatrix}_{i=1}^{M}$ to compute cluster centers $\begin{bmatrix} \boldsymbol{\beta}_i \end{bmatrix}_{l=1}^L$. The time segmentation decomposition is then given by
\begin{equation}
    \label{ts}
    e^{-j2\pi \phi(\mathbf{r}, t)} \approx \sum_{l=1}^L e^{-j2\pi \boldsymbol{\phi}(\mathbf{r}) \cdot \boldsymbol{\beta}_l} h_l(t)
\end{equation}
In the time segmentation model, there are many choices for the temporal basis functions $h_l(t)$. Zero-order interpolators are simple but will likely require a large $L$ for an accurate approximation, where as least squares interpolators will result in a smaller required $L$. Since the reconstruction time is directly proportional to $L$, it is desirable to use implicit GROG to further reduce the required number of segments $L$, or equivalently to improve the approximation given a fixed $L$. 

\subsection*{Implicit GROG with Field Correction}
To further improve the low-rank model, we propose to use GRAPPA kernels to model the multiplicative residual coil sensitivity weighted phase:
\begin{equation}
    \label{implicit_field}
    \bigg(e^{-j2\pi \phi(\mathbf{r}, t)} \bigg) \bigg( \mathbf{G}(t) \begin{bmatrix}
    \mathbf{s}(\mathbf{r}) e^{-j2\pi \mathbf{d}_1(t) \cdot \mathbf{r}} \\ \vdots  \\ \mathbf{s}(\mathbf{r}) e^{-j2\pi \mathbf{d}_D(t) \cdot \mathbf{r}}\end{bmatrix} \bigg) \approx \bigg( \sum_{l=1}^L e^{-j2\pi \boldsymbol{\phi}(\mathbf{r}) \cdot \boldsymbol{\beta}_l} h_l(t) \bigg) \mathbf{s}(\mathbf{r})
\end{equation}
Note that the GRAPPA basis functions $s(\mathbf{r})e^{-j2\pi \mathbf{d}_i \cdot \mathbf{r}}$ are time dependent, and hence the GRAPPA kernels must be estimated for each point along the k-space trajectory. As with the gridding case, the Implicit GROG framework can be used to instead estimate a low-dimensional representation of these GRAPAP kernels with an MLP. After applying the trained kernels $\mathbf{G}(t)$ to each group of $D$ samples, the reconstruction procedure is identical, with a significantly lower field correction fitting error due to the assistance from the GRAPPA kernels. This can be viewed as splitting the phase due to field imperfections at a given time point into a term that can be corrected by applying GRAPPA kernels to the raw data, and a term that must be corrected in the iterative reconstruction. 

\begin{figure}
    \centering
    \includegraphics[width=1\textwidth]{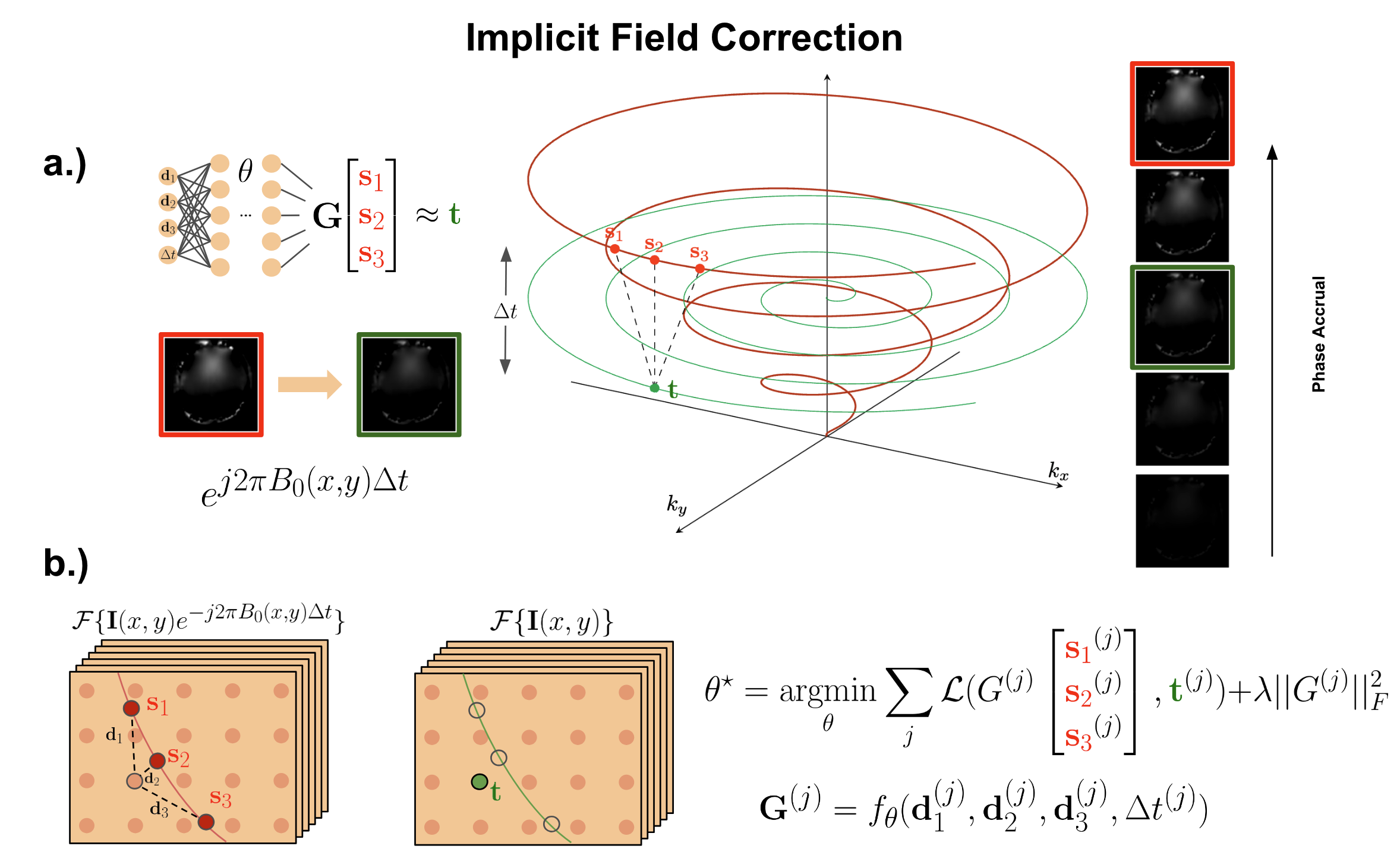}
    \caption{The Implicit GROG field correcting framework is shown for the special case of $\phi(\mathbf{r}, t) = B_0(\mathbf{r}) \cdot t$.  a.) The implicit neural network models the time varying GRAPPA kernels required to remove phase due to field imperfections. A unique kernel is needed for each point along the 2D spiral trajectory due to the time varying phase and orientation of the source points. b.) Training the implicit network is performed in a self supervised manner on multi-echo calibration data.}
    \label{fig:explain_b0}
\end{figure}

The GRAPPA kernels $\mathbf{G}(t)$ are now dependent on both orientation vectors $\mathbf{d}_i(t)$ and phase accrual $\phi(\mathbf{r}, t)$. If the field imperfection is linear in time (i.e. $B_0$ inhomogeneity), then using the elapsed time $\Delta t$ as an input to the MLP is a good choice, as shown in Figure \ref{fig:explain_b0}. For the more general case, it is desirable to instead use the features $[\boldsymbol{\alpha}(t) - \boldsymbol{\beta}_l]_{l=1}^L$ as input to the MLP.

\section*{Methods}
\label{sec:methods}

The implementation methodology for implicit GROG involves first training the GRAPPA Kernel MLP $f_\theta$, and then using this MLP on the raw data. 

\subsection*{GRAPPA Kernel Training}
In order to train the MLP $f_\theta(\*{d}_1, \cdots, \*{d}_{D}, \Delta t)$, we must provide a sufficient number of labeled training examples. We can generate these training examples from the calibration data. The MLP is trained according to the following procedure:
\begin{enumerate}
    \item Store all possible orientation vectors needed for gridding, \textit{or} estimate a distribution of orientation vectors closely resembling all possible orientation vectors needed for gridding
    \item Randomly choose a batch of $B$ target coordinates within the calibration bounds and use Kaiser Bessel interpolation to compute $B$ target values $\bigg\{ \mathbf{t}^{(j)} \hspace{.1cm} \big| \hspace{.1cm} j \in [B]\bigg\}$. 
    \item Randomly choose a batch of $B$ orientation vectors $\bigg\{ \big(\mathbf{d}_1^{(j)}, \cdots , \mathbf{d}_{N_s}^{(j)}\big) \hspace{.1cm} \big| \hspace{.1cm} j \in [B]\bigg\}$ from 1. 
    \item Use the orientation vectors and target coordinates to compute a set of $B \cdot N_s$ source coordinates and using Kaiser Bessel interpolation compute $B$ source values $\bigg\{\big(\mathbf{s}_1^{(j)}, \cdots, \mathbf{s}_{N_s}^{(j)} \big) \hspace{.1cm} \big| \hspace{.1cm} j \in [B]\bigg\}$
    \item Update MLP parameters $\theta$ by taking a gradient step in the direction of minimizing the training loss $\sum_{j=1}^B \mathcal{L}(f_\theta(\mathbf{d}_1^{(j)}, \cdots , \mathbf{d}_{N_s}^{(j)}) \begin{bmatrix} \mathbf{s}_1^{(j)} & \cdots & \mathbf{s}_{N_s}^{(j)} \end{bmatrix}^\top - \mathbf{t}^{(j)}) + \lambda ||f_\theta(\mathbf{d}_1^{(j)}, \cdots, \mathbf{d}_{N_s}^{(j)})||_F^2$
    \item Go back to 2. until convergence
\end{enumerate}

The MLP $f_\theta$ contains an input layer of size $d \cdot N_s$, a few hidden layers of size 256, and an output layer of size $N_s \cdot N_c \cdot N_c$ where $d$ is the spatial dimension, $N_s$ is the number of source points, and $N_c$ is the number of coils.

\subsection*{Applying the GRAPPA Kernels MLP}
Given a fully trained GRAPPA kernel MLP $f_\theta^\star$, we can estimate Cartesian data samples from non-Cartesian samples as the data is being acquired from the scanner. This can be done in real time, since the source points are chosen to be along the readout direction, and hence data from each phase encode can be gridded independently. As shown in Figure \ref{fig:explain}, we used a 1D interpolation to compute the samples $\mathbf{s}_1, \mathbf{s}_2, \mathbf{s}_3$. These interpolated source points are chosen along an arc with some constant arc-length, where the center sample having minimal distance to the Cartesian coordinate $\mathbf{t}$. Using data points along the readout direction only was done for simplicity of implementation. Once can also use fast non-Cartesian GRAPPA techniques \cite{luo2019grappa} to use source points from multiple phase encodes to estimate a single Cartesian sample.

\subsection*{Efficient Linear Operator Construction}
Both Gridded and non-Gridded forward models are implemented using PyTorch. The gridded forward model is implemented using simple FFT and masking operations. NUFFTs are implemented using the torchkbnufft package \cite{muckley2020torchkbnufft} with default parameters. 

The Topelitz method is used when possible for the non-Gridded reconstructions. The Toeplitz formulation has been extended to apply to subspace and time segmented reconstructions. Coil, subspace, and time-segment batching have been implemented in order to allow the Topelitz gram operator to fit into GPU memory. This extensive batching requires many CPU to GPU (and vice versa) memory writes, which significantly reduce reconstruction times, while still improving over NUFFTs. 

\subsection*{Reconstruction}
For all non-regularized reconstructions, the Conjugate Gradient algorithm is used to solve for the image. Regularized reconstructions use FISTA with $L_1$ Wavelet regularization. The regularization strength is empirically chosen by running multiple regularized reconstructions with different regularization strength, and choosing the best one. Before running the iterative reconstructions, the forward operators are normalized by $\frac{1}{\sqrt{\lambda_\text{max}}}$, where $\lambda_\text{max}$ is the maximum eigenvalue of the normal operator. All reconstructions, including gridded reconstructions, use the pipe-menon method to estimate the density compensation factor (DCF). It is important to note that the DCF for the gridded trajectory differs slightly from the DCF for the non-gridded trajectory. Estimated DCFs are normalized by the maximum DCF value.

\subsection*{iGROG and GROG Parameters}
Unless stated otherwise, Implicit GROG gridding uses a fixed MLP architecture and training scheme for all reconstructions. The MLP architecture has an input layer for the orientations, four hidden layers with 256 nodes, and the output layer with the same dimensions as the GRAPPA kernel. The ADAM \cite{kingma2014adam} optimizer is used with a learning rate of $10^{-3}$ to train the MLP under the L1 loss function $\mathcal{L}(\cdot) = ||\cdot||_1$. The model is trained over 5000 epochs with a batch size of 1024.

Coordinate GROG kernels have been trained using a set of different tikonov regularization values. For each experiment, the best reconstruction is chosen from the set of different regularization values.

\subsection*{Readout Interpolation}
Readout samples are interpolated in order to minimize the gridding distances (and hence gridding errors) for both iGROG and GROG, as shown in Figure \ref{fig:explain}. These interpolated samples are non-uniform relative to the raw temporal k-space signal. Sinc interpolation is the preferred choice for synthesizing these non-uniform interpolated samples. However, efficient algorithms are hard to come by due to the requirement of non-uniform interpolated samples, which disqualifies filter-bank or Fourier techniques. In order to have efficient readout interpolation, we first up-sample the raw data to a fixed 4X oversampling rate, and finally use a linear interpolation to estimate the non-uniform samples. This has been validated against a standard sinc interpolation, and achieves similar results.

\subsection*{Calibration}
Since both iGROG and GROG need a calibration region to train their respective GRAPPA kernel representations, a calibration region is either explicitly provided to the algorithm by a separate scan, or is synthesized from the center of k-space in the case of MRF subspace experiments. Calibration synthesis can be performed with MRF acquisitions due to the oversampling of the combined contrast center k-space. In order to ensure that edge effects do not corrupt the GRAPPA kernel estimation process, we omit a region of $2\Delta k$ points around the boundary of the calibration region.

\subsection*{MRF Subspace Reconstructions}
Magnetic Resonance Fingerprinting, or MRF, aims to target underlying tissue parameters from a time series of images \cite{ma2013magnetic}. More specifically, the image at each time point represents a new contrast that is evolving due to a train of varying flip angles. Each time point will have a unique contrast evolution that is of interest for the fingerprinting task downstream. Previous methods \cite{zhao2018improved} have shown that these time-series images are highly correlated, and if the flip angle train is carefully constructed, the images will belong to a low dimensional temporal subspace. This temporal subspace can be estimated from simulated signal evolutions offline. Thus, all MRF acquisitions will show the low dimensions subspace coefficient images \cite{cao2022optimized}. From these subspace coefficient images, one can extract the temporal signal evolution for any given voxel, and perform dictionary matching to estimate underlying tissue parameters.

\section*{Results}
\label{sec:results}

We provide source code and phantom experiments in: https://github.com/danielabrahamgit/igrog.

\subsection*{Dependence on Coil Sensitivity Maps}
A ground truth Shepp-logan phantom with image matrix size of (256, 256) is used to simulate multi-channel data from a 16-shot variable density spiral trajectory with under sampling factor $R = 2$. The Sensitivity maps were used from a 3T 48 channel head-only receiver array. In order to observe the dependence of iGROG and GROG on coil sensitivity maps, CG-SENSE reconstructions are performed using a varying number of virtual coils. These virtual coils were selected by applying coil compression \cite{buehrer2007array} to the 48-channel k-space data. NRMSE is compared against NUFFT-based reconstructions on gridded k-space data. 

\begin{figure}[hbt!]
    \centering
    \includegraphics[width=0.8\textwidth]{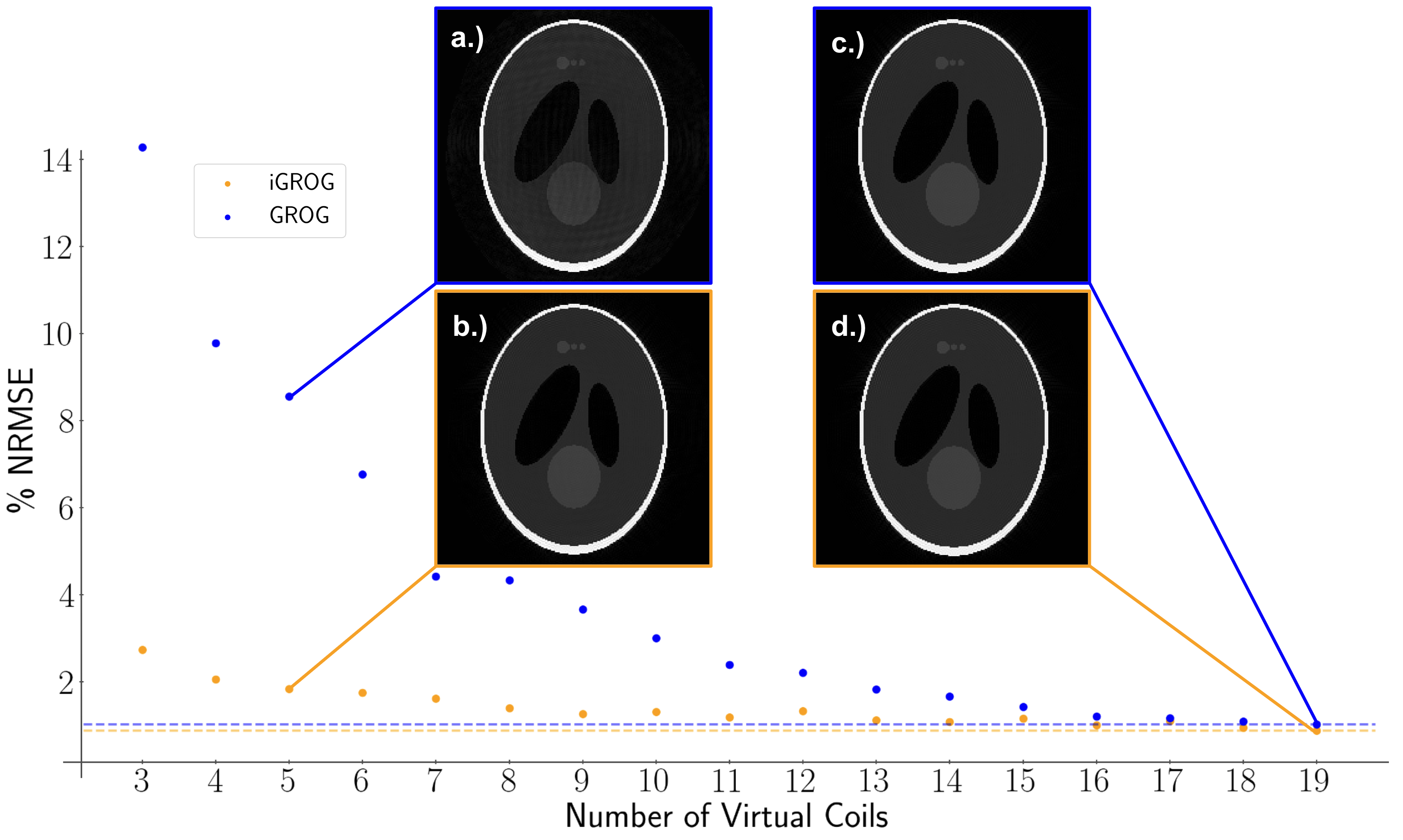}
    \caption{(a) GROG and (b) iGROG reconstructions from 5 virtual coil are shown along with (c) GROG and (d) iGROG reconstructions from 19 virtual coils.}
    \label{fig:coil_sweep}
\end{figure}

\subsection*{G-Factor Experiments}
To evaluate the bias and variance of Implicit GROG and GROG as k-space estimators, the pesudo-replica g-factor \cite{robson2008comprehensive} method is used. For each noise iteration, Implicit GROG and GROG are reconstructed for multiple values of $\lambda$, which controls the $L_2$ regularization on the GRAPPA kernels. The 'best' reconstructions for both methods are selected by finding a suitable trade off between noise amplification and artifact in the final image reconstruction. 

\begin{figure}[hbt!]
    \centering
    \includegraphics[width=0.8\textwidth]{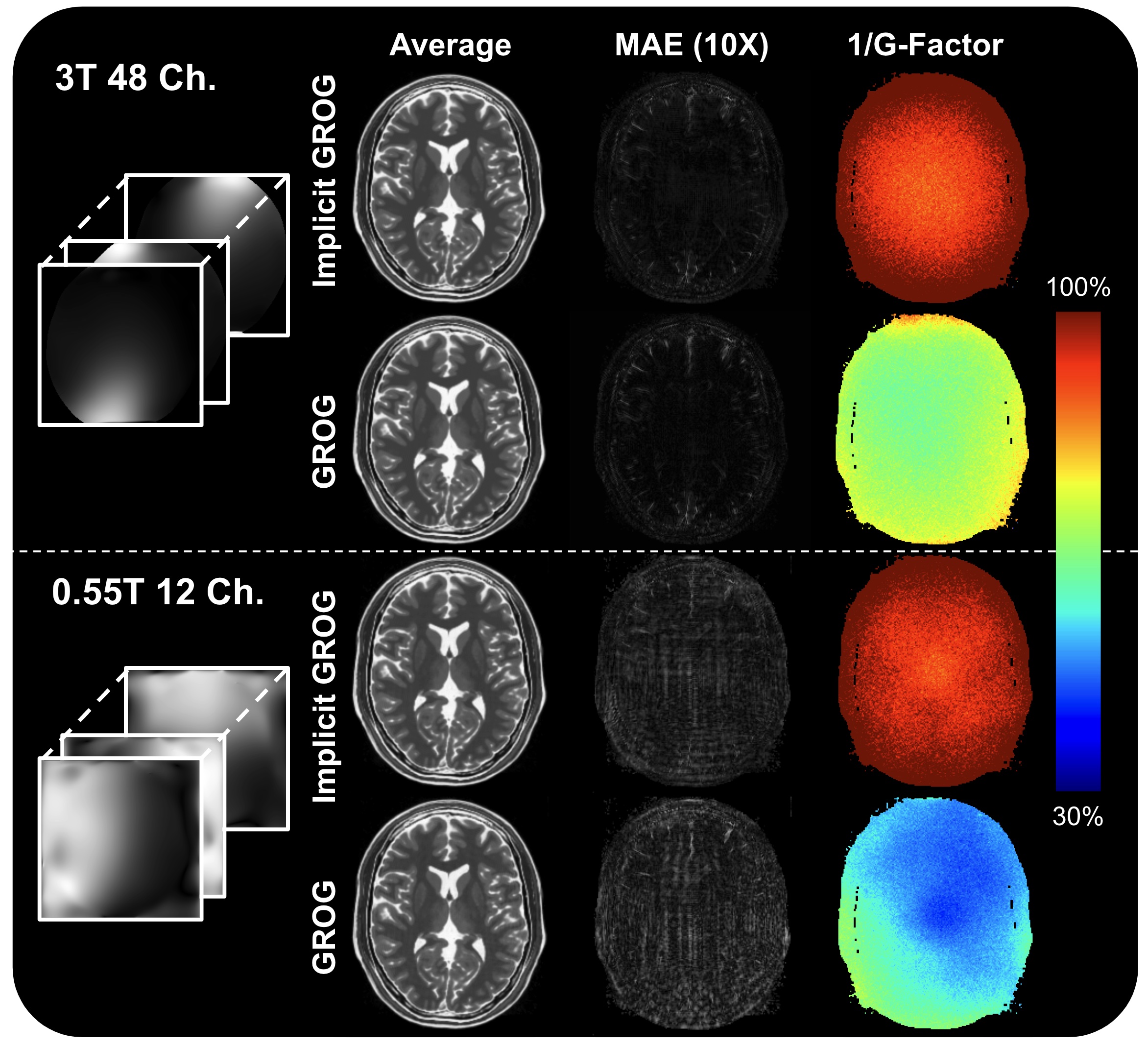}
    \caption{a.) SENSE reconstructions are repeated over 500 noise iterations to produce empirical G-factor (column 1) and bias (column 2) maps. This is repeated for two sets of sensitivity maps showcasing normal (top row) and poor (bottom row) sensitivity maps.}
    \label{fig:gfactor}
\end{figure}

\subsection*{2D MRF}
MRF acquisitions aim to resolve temporal signal evaluations in order to resolve underlying 
A 16-shot variable density spiral trajectory is used to cover the 2D k-space over 600 time points. Each time point will contain a single interleave of the 16-shot variable density spiral. The flip angle train used is described in \cite{cao2022optimized}, and is optimized for separating signal signatures from grey and white matter. Temporal subspace coefficient images are reconstructed using CG-SENSE. The raw 48-channel data was compressed down to 12 channels using coil compression.

\begin{figure}[hbt!]
    \centering
    \includegraphics[width=0.8\textwidth]{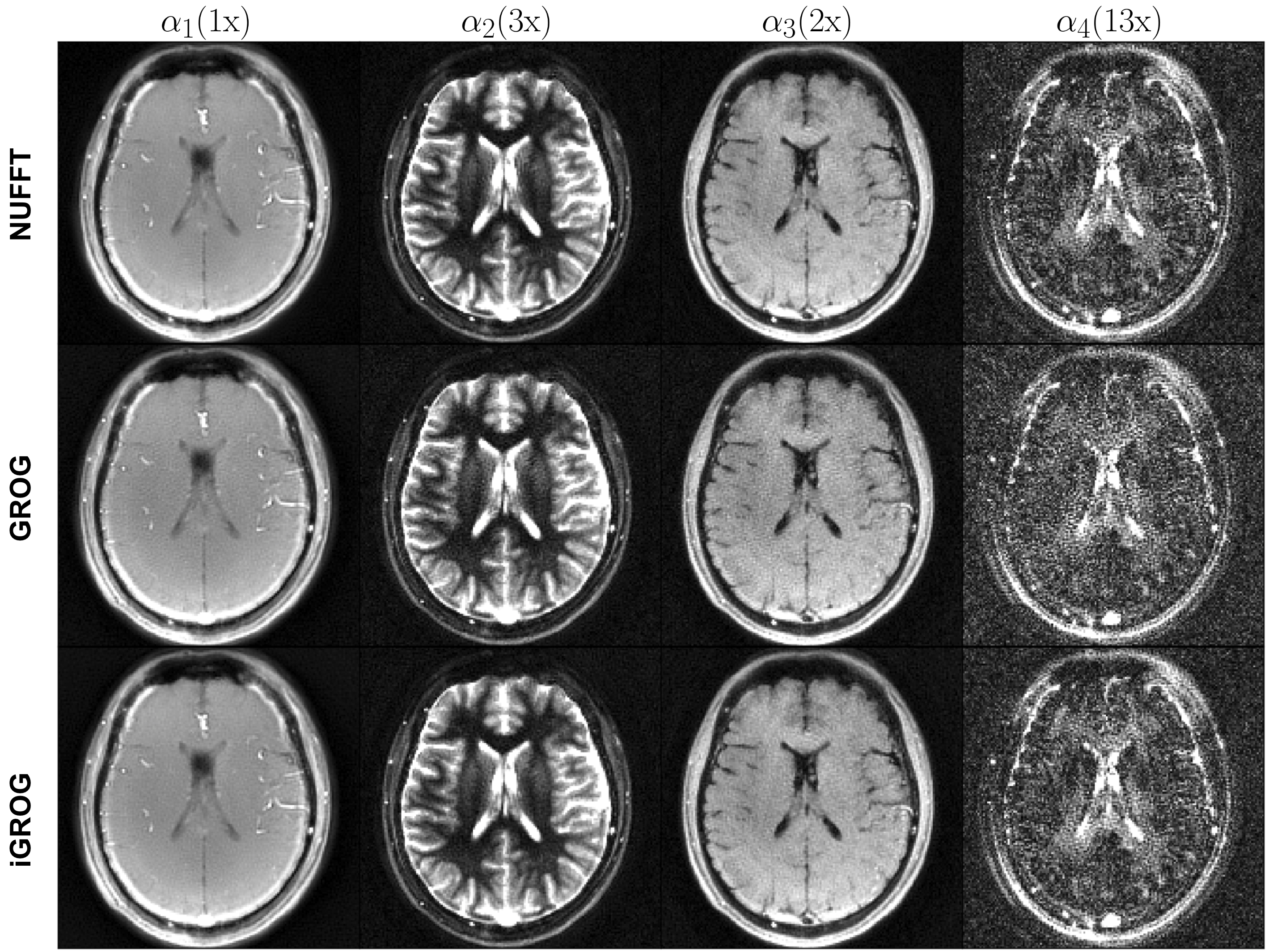}
    \caption{CG-SENSE 2D MRF Subspace Coefficients. The rows show the NUFFT-based, iGROG, and GROG reconstructions respectively. The columns show the different subspace coefficients.}
    \label{fig:mrf_2d}
\end{figure}

\subsection*{3D MRF At 0.5T}
The 3D MRF data acquisiton was performed on a lowfield 0.5T scanner. At low field, the coil sensitivity maps are much less effective at spatial encoding. For this reason, we use iGROG and GROG to grid to a 1.2X oversampling grid. While not as fast as a 1X grid, this will still have a much more efficient reconstruction in comparison to NUFFTs. To help combat the low SNR setting of lowield scanners, we use a locally low rank (LLR) regularized subspace reconstruction. This regularization is realized via the FISTA \cite{beck2009fast} algorithm. The locally low rank blocks are of size (10, 10, 10) and have a stride of (10, 10, 10). The LLR regularization parameter was chosen by running the standard NUFFT based reconstruction over multiple values, and selecting the reconstruction that best trades off noise and block artifacts. For iGROG and GROG reconstructions, the LLR regularization parameter was chosen to minimize the error with respect to the NUFFT reconstruction. This must be done since both the data and forward model are augmented in gridded reconstructions.

Calculating $A^H\mathbf{b}$ using NUFFT took 5.8min, while iGROG and GROG took about 3 seconds to calculate $A_\text{grid}^H \mathbf{b}$. Calculating 40 FISTA iterations using the standard Toeplitz formulation took 16min, while 40 FISTA iterations with iGROG/GROG took 4.3min. 

\begin{figure}[hbt!]
    \centering
    \includegraphics[width=1.0\textwidth]{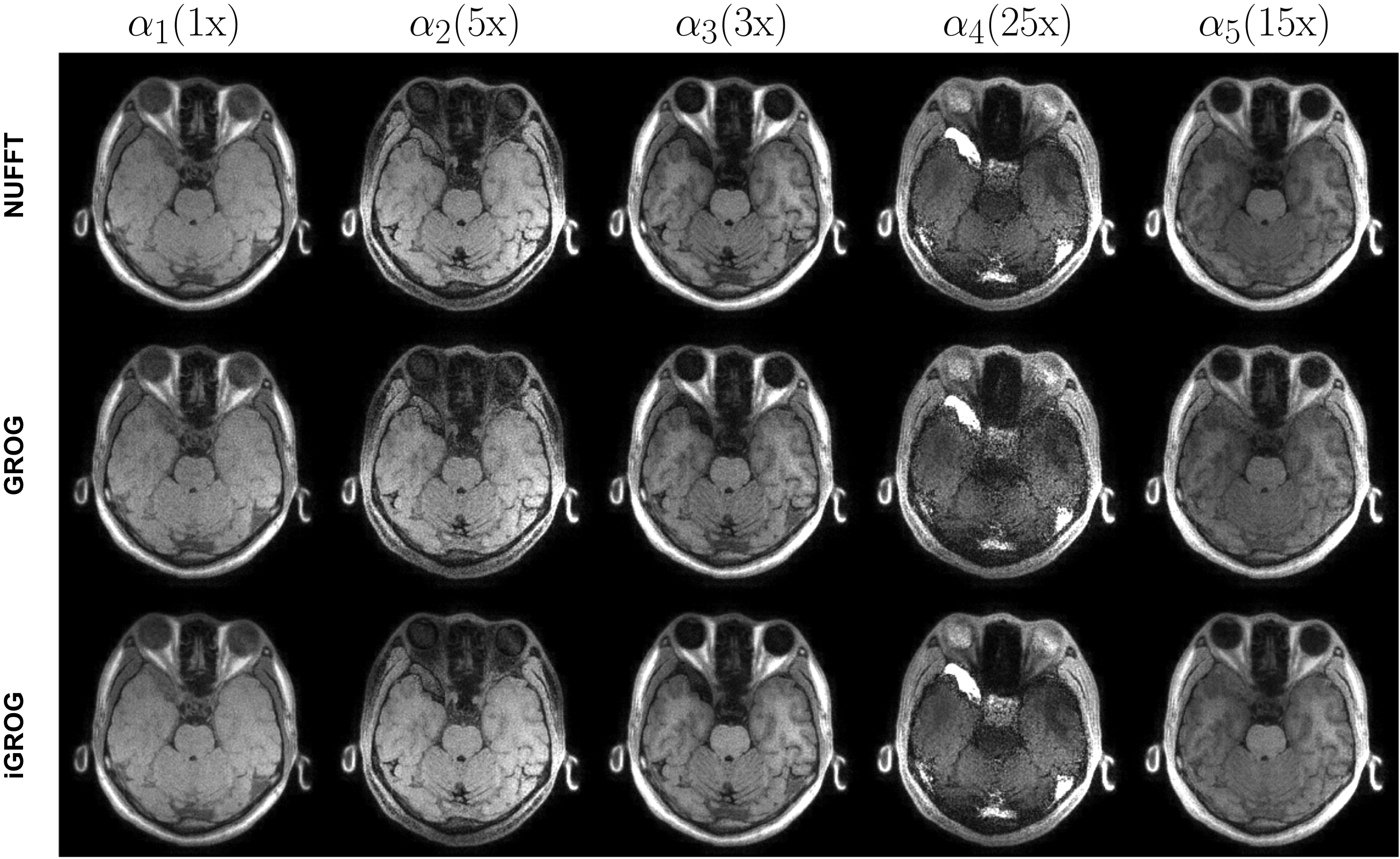}
    \caption{CG-SENSE 3D Low-field MRF Subspace Coefficients}
    \label{fig:mrf_3d}
\end{figure}

\subsection*{MRF at 3T}
MRF reconstructions are also performed on a 20 channel 3T scanner with and without Implicit GROG to demonstrate timing differences.

\begin{figure}[hbt!]
    \centering
    \includegraphics[width=1.0\textwidth]{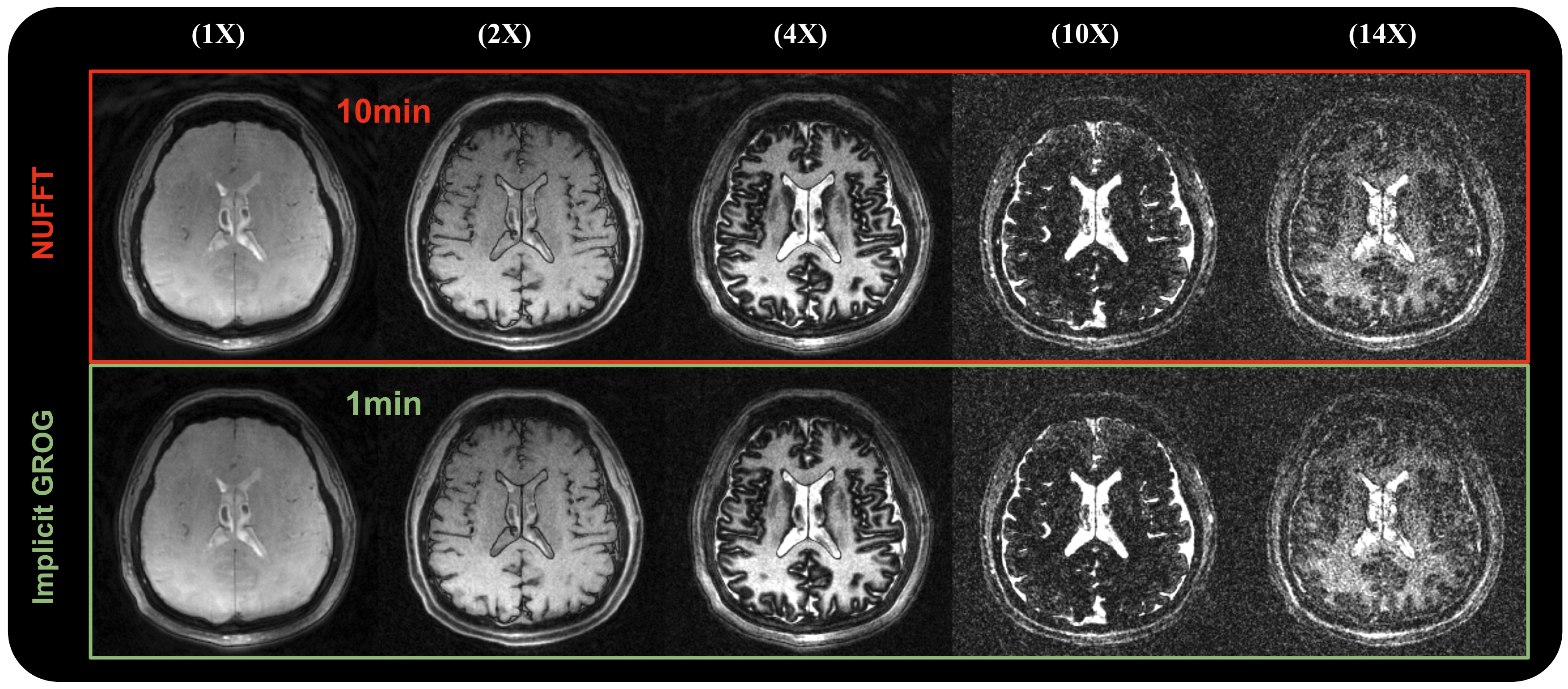}
    \caption{MRF data from a 3T 20 channel head-coil is reconstructed using CG-SENSE into subspace coefficients representing time varying contrast changes. First row shows the standard NUFFT-based reconstruction, the second row shows implicit GROG reconstructions. Both have }
    \label{fig:20ch_mrf}
\end{figure}

\subsection*{Time Segmentation Simulation}
A 32-channel 3T GE head coil is used to simulate a 4-shot 60ms spiral trajectory from a T1 weighted brain phantom. Time segmentation is performed on both reconstructions, with and without implicit field correction. A simple zero-order hold function is used for the temporal interpolation in the time segmentation algorithm. 

\begin{figure}[hbt!]
    \centering
    \includegraphics[width=1.0\textwidth]{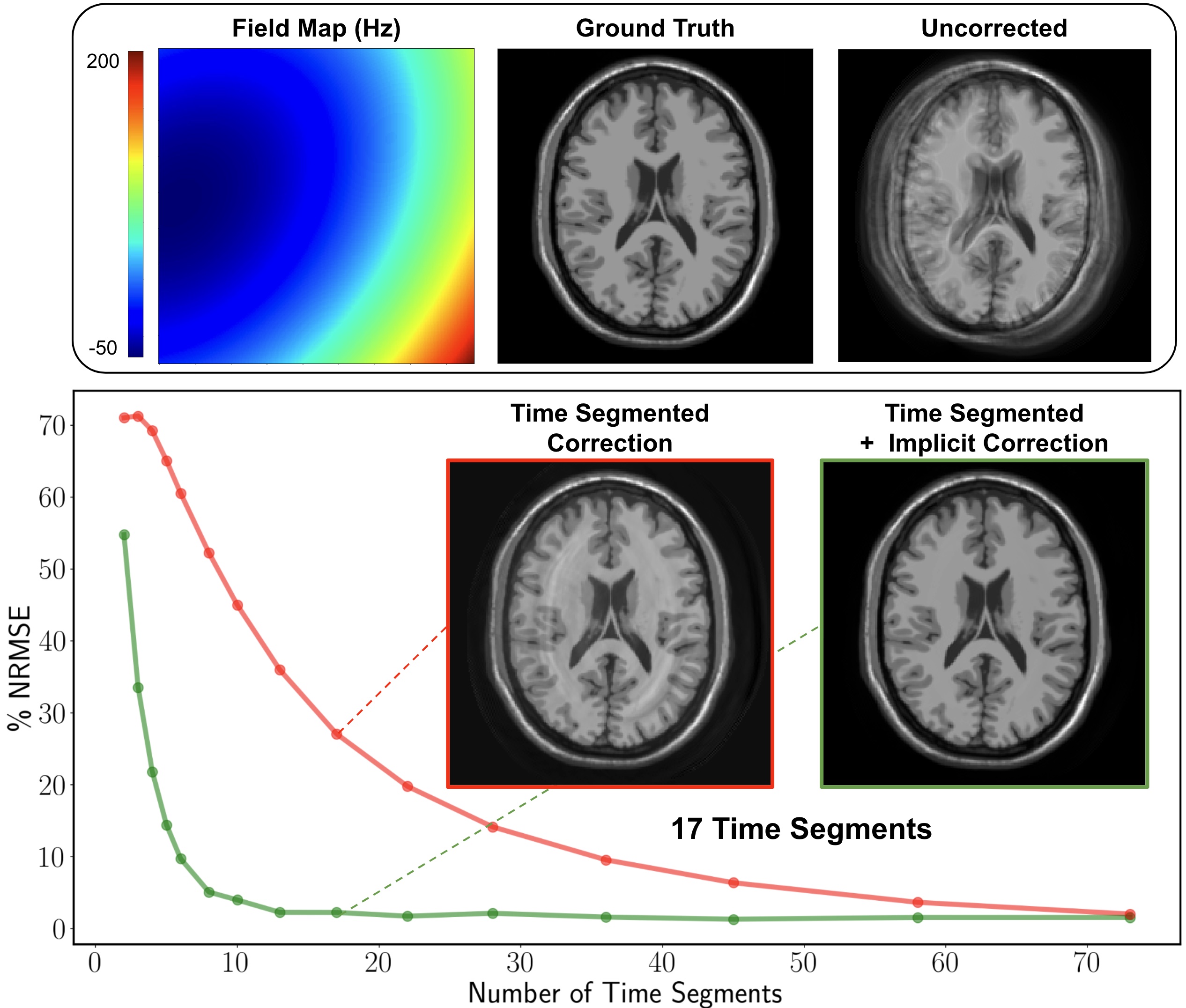}
    \caption{Top row shows a (left) simulated second order field map applied to a (middle) brain phantom (right) the artifact. Bottom plot shows (red) standard time segmented correction over multiple segments and time segmentation with (green) implicit field correction. Reconstructions with 17 segments are shown.}
    \label{fig:time_seg}
\end{figure}

\section*{Discussion}
Experiments confirm that gridded reconstructions indeed provide significant computational improvements in comparison to NUFFT-based reconstructions. For iterative methods that repeatedly evaluate the normal operator, the use of the Toeplitz method is currently severely limited by GPU memory transfers. Gridded reconstructions do not suffer from this problem, and hence have faster reconstructions due to both algorithmic benefits and larger allowable GPU batches.

While both Implicit GROG and GROG provide identical copmutational improvements, important differences between the two algorithms result in different reconstruction quality. Simulation experiments in Figure \ref{fig:coil_sweep} show that GROG requires more coil information in order to achieve the same RMSE as Implicit GROG does. Given a sufficient set of coil sensitivity maps, Figure \ref{fig:gfactor} shows that GROG can produce reasonable reconstructions with an SNR penalty of roughly 30\%, while implicit GROG has an SNR penalty closer to 5\%. This suggests that implicit GROG can be used as a suitable alternative for NUFFT-based reconstructions in cases where the a gridded trajectory PSF is acceptable. 

The most advantageous feature about implicit representations of GRAPPA kernels is the ability to train a vast number of unique GRAPPA kernels in a compressed way. As shown, this provides desirable computational improvements when using such kernels to model off grid sampling and field imperfections. It is important to note that all methods and results have been built upon selecting source points from the readout direction only. This can be extended to incorporate a general selection of source and target points, and can even be used for non-Cartesian GRAPPA reconstructions \cite{luo2019grappa}.

\section*{Conclusion}
\label{sec:conclusion}

In this work, we provide a generalized framework for rapid non-Cartesian reconstruction by extending the previous GROG technique to properly use multiple source points, and allowing for modeling of field imperfections. An implicit MLP is used to compactly and implicitly represent a large collection of GRAPPA kernels. By removing translating the field corrupted non-Cartesian data to only partially corrupted Cartesian data, the new gridded forward model has a reduced computatinoal complexity, enabling much faster reconstruction times.

\bibliography{main}

\begin{thebibliography}{10}

\bibitem{feng2016xd}
Feng~L, Axel~L, Chandarana~H, Block~KT, Sodickson~DK, Otazo~R.
\newblock Xd-grasp: golden-angle radial mri with reconstruction of extra
  motion-state dimensions using compressed sensing.
\newblock Magnetic resonance in medicine 2016; 75:775--788.

\bibitem{Wiesinger2018}
Wiesinger~F, Menini~A, Solana~AB.
\newblock Looping star.
\newblock Magnetic Resonance in Medicine 2018; 81:57--68.

\bibitem{fessler2003nonuniform}
Fessler~JA, Sutton~BP.
\newblock Nonuniform fast fourier transforms using min-max interpolation.
\newblock IEEE transactions on signal processing 2003; 51:560--574.

\bibitem{baron2018rapid}
Baron~CA, Dwork~N, Pauly~JM, Nishimura~DG.
\newblock Rapid compressed sensing reconstruction of 3d non-cartesian mri.
\newblock Magnetic resonance in medicine 2018; 79:2685--2692.

\bibitem{fessler2005toeplitz}
Fessler~JA, Lee~S, Olafsson~VT, Shi~HR, Noll~DC.
\newblock Toeplitz-based iterative image reconstruction for mri with correction
  for magnetic field inhomogeneity.
\newblock IEEE Transactions on Signal Processing 2005; 53:3393--3402.

\bibitem{man1997multifrequency}
Man~LC, Pauly~JM, Macovski~A.
\newblock Multifrequency interpolation for fast off-resonance correction.
\newblock Magnetic resonance in medicine 1997; 37:785--792.

\bibitem{sutton2003fast}
Sutton~BP, Noll~DC, Fessler~JA.
\newblock Fast, iterative image reconstruction for mri in the presence of field
  inhomogeneities.
\newblock IEEE transactions on medical imaging 2003; 22:178--188.

\bibitem{seiberlich2007non}
Seiberlich~N, Breuer~FA, Blaimer~M, Barkauskas~K, Jakob~PM, Griswold~MA.
\newblock Non-cartesian data reconstruction using grappa operator gridding
  (grog).
\newblock Magnetic Resonance in Medicine: An Official Journal of the
  International Society for Magnetic Resonance in Medicine 2007; 58:1257--1265.

\bibitem{griswold2002generalized}
Griswold~MA, Jakob~PM, Heidemann~RM, Nittka~M, Jellus~V, Wang~J, Kiefer~B,
  Haase~A.
\newblock Generalized autocalibrating partially parallel acquisitions (grappa).
\newblock Magnetic Resonance in Medicine: An Official Journal of the
  International Society for Magnetic Resonance in Medicine 2002; 47:1202--1210.

\bibitem{luo2019grappa}
Luo~T, Noll~DC, Fessler~JA, Nielsen~JF.
\newblock A grappa algorithm for arbitrary 2d/3d non-cartesian sampling
  trajectories with rapid calibration.
\newblock Magnetic resonance in medicine 2019; 82:1101--1112.

\bibitem{mildenhall2021nerf}
Mildenhall~B, Srinivasan~PP, Tancik~M, Barron~JT, Ramamoorthi~R, Ng~R.
\newblock Nerf: Representing scenes as neural radiance fields for view
  synthesis.
\newblock Communications of the ACM 2021; 65:99--106.

\bibitem{muckley2020torchkbnufft}
Muckley~MJ, Stern~R, Murrell~T, Knoll~F.
\newblock Torchkbnufft: A high-level, hardware-agnostic non-uniform fast
  fourier transform.
\newblock ISMRM Workshop on Data Sampling \& Image Reconstruction 2020; p.~22.

\bibitem{kingma2014adam}
Kingma~DP, Ba~J.
\newblock Adam: A method for stochastic optimization.
\newblock arXiv preprint arXiv:1412.6980 2014; .

\bibitem{ma2013magnetic}
Ma~D, Gulani~V, Seiberlich~N, Liu~K, Sunshine~JL, Duerk~JL, Griswold~MA.
\newblock Magnetic resonance fingerprinting.
\newblock Nature 2013; 495:187--192.

\bibitem{zhao2018improved}
Zhao~B, Setsompop~K, Adalsteinsson~E, Gagoski~B, Ye~H, Ma~D, Jiang~Y,
  EllenGrant~P, Griswold~MA, Wald~LL.
\newblock Improved magnetic resonance fingerprinting reconstruction with
  low-rank and subspace modeling.
\newblock Magnetic resonance in medicine 2018; 79:933--942.

\bibitem{cao2022optimized}
Cao~X, Liao~C, Iyer~SS, Wang~Z, Zhou~Z, Dai~E, Liberman~G, Dong~Z, Gong~T, He~H
  et~al.
\newblock Optimized multi-axis spiral projection mr fingerprinting with
  subspace reconstruction for rapid whole-brain high-isotropic-resolution
  quantitative imaging.
\newblock Magnetic Resonance in Medicine 2022; 88:133--150.

\bibitem{buehrer2007array}
Buehrer~M, Pruessmann~KP, Boesiger~P, Kozerke~S.
\newblock Array compression for mri with large coil arrays.
\newblock Magnetic Resonance in Medicine: An Official Journal of the
  International Society for Magnetic Resonance in Medicine 2007; 57:1131--1139.

\bibitem{robson2008comprehensive}
Robson~PM, Grant~AK, Madhuranthakam~AJ, Lattanzi~R, Sodickson~DK, McKenzie~CA.
\newblock Comprehensive quantification of signal-to-noise ratio and g-factor
  for image-based and k-space-based parallel imaging reconstructions.
\newblock Magnetic Resonance in Medicine: An Official Journal of the
  International Society for Magnetic Resonance in Medicine 2008; 60:895--907.

\bibitem{beck2009fast}
Beck~A, Teboulle~M.
\newblock A fast iterative shrinkage-thresholding algorithm for linear inverse
  problems.
\newblock SIAM journal on imaging sciences 2009; 2:183--202.

\end{thebibliography}

\clearpage

\listoffigures

\end{document}